\documentclass{ws-p8-50x6-00}

\newcommand{\vslash}{\mbox{$\not{\hspace{-1.03mm}v}$}}        
\newcommand{\zslash}{\mbox{$\not{\hspace{-1.03mm}z}$}}
\newcommand{\nslash}{\mbox{$\not{\hspace{-1.03mm}n}$}}

\begin{document}

\title{Rare radiative B decays in perturbative QCD}

\author{Dan Pirjol}

\address{Department of Physics, University of California
at San Diego\\
9500 Gilman Drive, La Jolla, CA 92093}


\maketitle

\abstracts{
We report on recent progress on perturbative QCD calculations
of certain exclusive rare weak $B$ meson decays involving hard photons.
In the limit of a photon energy $E_\gamma$ much larger than 
$\Lambda_{QCD}$, the amplitudes for such processes can be 
analyzed in a twist expansion in powers of $\Lambda/E_\gamma$. The 
leading twist amplitude is given by the convolution of a hard
scattering amplitude with the $B$ meson light-cone wavefunction.
This approach is applied to a calculation of the leptonic radiative 
$B\to \gamma \ell \nu_\ell$ formfactors and to an estimate
of the weak annihilation contribution to the penguin decays $B\to
\rho(\omega)\gamma$. As an application we discuss a few methods for
constraining the unitarity triangle with exclusive radiative $B$
decays.
}

\section{Introduction}

Rare radiative decays of $B$ hadrons have been extensively studied as 
a possible way of probing possible new physics effects. Most 
effort has been concentrated on the theoretically cleaner inclusive 
decays, which can be analyzed in an $1/m_b$ expansion
with the help of an operator product expansion. 

On the other hand, exclusive rare $B$ decays are sensitive
to long-distance QCD effects and the relevant amplitudes depend on
details of the hadronic bound states. This makes their theoretical
description considerably more model-dependent, and a full description
is still lacking.

In the following we describe a systematic treatment of exclusive
decays involving one very energetic on-shell photon emitted from an
internal quark line. In this situation the internal light
quark moves very close to the light cone, and an expansion in powers
of the small parameter $\Lambda/E_\gamma$ becomes 
useful\cite{ope}. This is
analogous to the twist expansion for exclusive processes involving
only light quarks\cite{BroLe}.

The simplest process which can be studied along these lines is the 
leptonic radiative decay\cite{KPY} $B\to \gamma\ell \nu_\ell$, which is
discussed in Sec.~2. A basic ingredient of the method is the 
light-cone description of a heavy-light meson, which is introduced 
in Sec.~2.1. In Sec.~2.2 we argue that
the general structure of the leading-twist $B\to \gamma\ell \nu_\ell$ 
form factors is given by a convolution of the $B$ light-cone
wave function with a hard scattering amplitude $T_H$.

In Sec.~3 we describe an application of this method to the
calculation\cite{GP}
of the weak annihilation contribution to the weak radiative
decay $B\to \rho\gamma$. 
Taking into account all possible long distance amplitudes 
compatible with SU(3) symmetry, we present in Sec.~3.1 a few methods
for constraining the unitarity triangle using
isospin violating effects in $B\to \rho(\omega)\gamma$ decays\cite{bound}.

\section{Leading twist calculation of the $B\to\gamma \ell\nu_\ell$
decay}

\subsection{Heavy mesons light-cone wavefunctions}

The most general form for the Bethe-Salpeter wave function of a
heavy-light meson $B(v)$ can be written in terms of 4 scalar functions
$\psi_i(v\cdot z,z^2)$ ($z=x-y$)
\begin{equation}\label{1}
\psi_{\alpha\beta}(z,v) = \langle 0|Tb_\alpha (x) \bar q_\beta(y)|
\bar B(v)\rangle =
\left\{ (
\psi_1 - \frac12 (\zslash \vslash - \vslash \zslash )\psi_2
+ \vslash\psi_3 + \zslash \psi_4 )\right\}_{\alpha\beta}\,.
\end{equation}
In the heavy quark limit the structure of the wavefunction
is further constrained by the condition $\vslash\psi = \psi$. This
reduces the number of independent structures to 2, which can be
taken as $\psi_{1,2}$ 
\begin{equation}\label{2}
\psi_{\alpha\beta}(z,v) = \left\{\frac{1+\vslash}{2}
(\psi_1 + [\zslash - \vslash (v\cdot z) ]\psi_2 )
\gamma_5\right\}_{\alpha\beta}\,.
\end{equation}
The momentum space $B$ meson wavefunction is defined as
\bea\label{3}
\psi_{\alpha\beta}(k) \equiv \int d^4 z e^{ik\cdot
z}\psi_{\alpha\beta}(v,z)\,.
\eea
In the physical applications to be discussed in the following, 
this wavefunction is convoluted with a scattering amplitude
$T_H(k)$ which
does not depend on one light-cone component of the relative momentum 
$k_\mu$ in the bound state (e.g. $k_-$). Therefore, this
component can be integrated over, which effectively puts $z$ on the 
light cone ($z_+=0$). We defined
here light-cone coordinates as $k_\pm \equiv k^0\pm k^3$, which can
be projected out by dotting into the light-cone basis vectors
$n=(1,0,0,1)$ and $\bar n=(1,0,0,-1)$. This gives $k_+=n\cdot k, 
k_- = \bar n\cdot k$.

It is convenient to write the wavefunction (\ref{2}) as a
linear combination
of light-cone projectors in spinor space $\Lambda_+ = \frac14
\nslash\overline{\nslash}\,,
\Lambda_- = \frac14 \overline{\nslash}\nslash$
\begin{equation}\label{4}
\psi_{\alpha\beta}(z_+=0,z_-,z_\perp = 0) = \left\{\frac{1+\vslash}{2}
(\Lambda_+ \psi_+(z_-) + \Lambda_- \psi_-(z_-))
\gamma_5\right\}_{\alpha\beta}\,.
\end{equation}
with $\psi_\pm(z_-) = \psi_1 \mp \frac12 z_-\psi_2$. [When used 
to compute a physical amplitude, only one of these functions
will contribute to leading twist (e.g. $\psi_+$).] 
This gives the most general expression
for the $B$ wavefunction in the heavy quark limit, which is
often found in the literature \cite{BBNS2,BeFe}
\begin{equation}\label{5}
\psi_{\alpha\beta}(v\cdot z,z^2) = \frac{1+\vslash}{2}
\left\{(\psi_+ - \frac{\zslash}{2v\cdot z}
(\psi_+ - \psi_- ))
\gamma_5\right\}_{\alpha\beta}\,.
\end{equation}
However, in the following we will use the form (\ref{4}) which has
a simple interpretation in the parton model. With the kinematics
adopted above, the wavefunction $\psi_+(k_+)$ gives the probability
to find the light quark in the $B$ meson carrying momentum $k_+$.
Its moments are related to matrix elements of local operators
\begin{equation}
\langle k^N_+ \rangle \equiv
\int_0^\infty dk_+ k_+^N \psi_+(k_+)  = 
\langle 0|\bar q \Lambda_+ \nslash
\gamma_5 (in\cdot D)^N h_v|\bar B(v)\rangle\,.
\end{equation}

The first two moments can be expressed in terms of known 
$B$ hadronic parameters: $\langle k^0_+\rangle = 
f_B m_B$
and $\langle k_+\rangle =  \frac43
\bar\Lambda f_B m_B$,
with $\bar \Lambda = m_B-m_b \simeq 350$ MeV the binding energy of the
$b$ quark in a $B$ meson. The average of $k^{-1}_+$ will play an
important role in the following. Although it cannot be related to the
matrix element of a local operator, it is possible to give a
model-independent lower bound on its magnitude.
Using the normalization conditions
for $N=0,1$ and the positivity of the distribution function $\psi_+$
one finds\cite{KPY}
\begin{equation}\label{bound}
\int_0^\infty dk_+ \frac{\psi_+(k_+)}{k_+} \geq 
\frac{3}{4\bar\Lambda} f_B m_B\,.
\end{equation}
For more specific (but less model-independent) predictions,
some model has to be
adopted for the distribution function $\psi_+(k_+)$. We used 
an ansatz inspired by a quark model with harmonic oscillator
potential $\psi_+(k_+)  = Nk_+
\exp\left(-\frac{1}{2\omega^2}(k_+-a)^2\right)$. The width parameter
$\omega$ is varied as $\omega = 0.1-0.3$ MeV and $a$ is constrained 
from the normalization conditions. Using $\bar\Lambda = 0.3-0.4$ MeV
gives $a=0.05-0.5$ MeV.

\subsection{Leading twist form factors in $B\to\gamma \ell \nu_\ell$}

\begin{figure}[t]
\centerline{
\mbox{\epsfysize=3.0truecm \hbox{\epsfbox{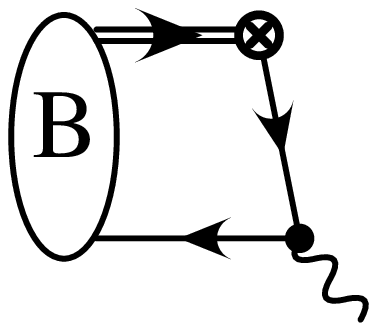}}}\qquad\qquad
\mbox{\epsfysize=3.0truecm\raise21pt \hbox{\epsfbox{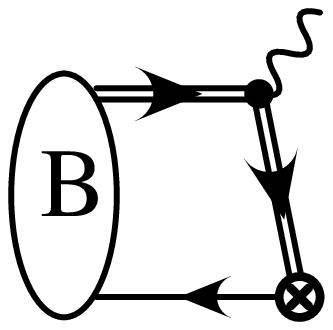}}}
}
\centerline{\mbox{(a)}\hspace{4.5cm} \mbox{(b)}}
\caption{Tree level contributions to the hard scattering
amplitude for $B\to \gamma \ell\nu_\ell$ decay.
The cross denotes the weak current
$J_\mu = \bar q\gamma_\mu (1-\gamma_5) b$. The dominant contribution
comes from the photon being emitted from the light quark line (a)
while the diagram (b) is suppressed by the inverse heavy quark mass
$1/m_b$.
\label{fig:tree}}
\end{figure}

The simplest hard photon process which can be analyzed using an 
expansion in $1/E_\gamma$ is the radiative leptonic decay 
$B\to\gamma \ell \nu_\ell$. This proceeds through the weak annihilation
of the spectator quark in the $B$ meson, as depicted in the quark
diagrams in Fig.~\ref{fig:tree}.

Interest in this decay was sparked by the observation\cite{BuGoWy} 
that its branching ratio is enhanced relative to that for the 
leptonic decay $B\to \ell\nu_\ell$ with $\ell=e,\mu$.
Although adding one photon introduces a factor of
$\alpha=\frac{1}{137}$
in the rate, it also removes the helicity suppression factor
$(m_\ell/m_B)^2$; the overall effect is an enhancement in the
rate of the leptonic radiative mode. Later work\cite{1} investigated 
this decay using a variety of approaches, both in connection with 
the prospect of constraining $f_B$ and/or $|V_{ub}|$ and in the 
context of the weak radiative decays $B\to \rho(\omega)\gamma$.

The amplitude for $B\to\gamma\ell\nu_\ell$ is parameterized by two 
formfactors $f_{V,A}(E_\gamma)$  defined as
\begin{eqnarray}\label{fV}
& &\frac{1}{\sqrt{4\pi\alpha}} \langle \gamma(q,\varepsilon)|
\bar q\gamma_\mu b|\bar B(v)\rangle =
i\epsilon_{\mu\alpha\beta\delta} \varepsilon^{*\alpha} v^\beta
q^\delta f_V(E_\gamma)\\ 
\label{fA}
& &\frac{1}{\sqrt{4\pi\alpha}} \langle \gamma(q,\varepsilon)|
\bar q\gamma_\mu\gamma_5 b|\bar B(v)\rangle =
\left[ q_\mu (v\cdot \varepsilon^*) - \varepsilon^*_\mu (v\cdot q) 
\right]  f_A(E_\gamma)\,.
\end{eqnarray}
Computing the contribution of the two diagrams in Fig.~\ref{fig:tree}
with the $B$ wavefunction (\ref{4}) one finds for the formfactors
(\ref{fV}), (\ref{fA}) 
\begin{equation}\label{fVA}
f_V(E_\gamma) = f_A(E_\gamma) = \frac{f_B m_B}{2E_\gamma}
\left(Q_q R - \frac{Q_b}{m_b}\right) + {\cal O}(\Lambda^2/E_\gamma^2)
\end{equation}
where $Q_q,Q_b$ are the spectator- and heavy-quark electric charges and
the hadronic parameter $R$ describes the photon coupling to the
light quark. This is given by a convolution of the leading twist 
$B$ meson wavefunction with a hard-scattering amplitude
$T_H(E_\gamma,k_+)$
\begin{equation}\label{R}
R(E_\gamma) = \int_0^\infty 
d k_+ \frac{\psi_+(k_+)}{k_+} T_H(E_\gamma,k_+)
\end{equation}
The hard scattering amplitude $T_H(E_\gamma,k_+)$ is given to one-loop
order by\cite{KPY}
\begin{equation}\label{TH}
T_H(E_\gamma,k_+) = 1 + \frac{\alpha_s C_F}{4\pi}\left(
-\log^2\left(\frac{2E_\gamma}{k_+}\right) + 
\frac52\log\left(\frac{2E_\gamma}{k_+}\right)
-\frac{4\pi^2}{3}\right)\,.
\end{equation}
Using the model distribution function described in Sec.~2.1 one finds
at tree level $R(E_\gamma)=2-4$ GeV$^{-1}$, while the lower bound 
(\ref{bound}) predicts $R > 2.1$ GeV$^{-1}$ (corresponding to 
$\bar\Lambda = 350$ MeV).

We briefly discuss in the following a few interesting consequences 
of these results.

a) To leading twist the formfactors $f_{V,A}(E_\gamma)$ scale
like $1/E_\gamma$, with a proportionality coefficient which is independent
on the heavy quark flavor (up to $1/m_b$ corrections). This property can be
used to determine the CKM matrix element $|V_{ub}|$ from a comparison
of the $B\to \gamma \ell \nu_\ell$ and $D\to \gamma \ell \nu_\ell$ photon
spectra\cite{KPY}.

b) The form factors of the vector and axial current are equal to 
leading twist. This is in contrast to their behaviour in the low $E_\gamma$
region, where they receive contributions from intermediate states with 
different quantum numbers\cite{BuGoWy} ($J^P=1^-$ for $f_V$ and 
$J^P=1^+$ for $f_A$). 

The equality of the form factors (\ref{fVA}) is a particular case of 
a symmetry 
relation analogous to those discussed for the soft components of
semileptonic formfactors in\cite{french}. The gluon couplings of a
quark moving close to the light cone 
possess a higher symmetry\cite{DuGri}. This can
be formalized by going over to an effective theory, which should
include in addition to the soft gluon modes (LEET\cite{DuGri}), also
collinear gluons. A complete discussion including collinear gluons 
has been given only recently\cite{BFPS}.

There is however an important difference between the status of the 
symmetry relation $f_V=f_A$ among the leptonic radiative formfactors, 
and the
analogous symmetry relations for the soft semileptonic
formfactors\cite{french}. While the latter receive symmetry breaking
corrections from hard gluon exchange (which have been computed
recently\cite{BeFe}), the former relation is not changed by such
effects, as checked by explicit calculation to one-loop
order\cite{KPY}.


Hadronic amplitudes similar to $R$ in (\ref{R}) appear in many physical
quantities involving long-distance effects induced by hard photon
or gluon emission from weak annihilation topologies\cite{BDS}.
[When computed in the quark model, this quantity appears as the
inverse of the constituent quark mass $R\to 1/m_q$.]
In the following section we study such an important application, 
to long-distance effects in exclusive penguin induced decays 
$b\to d\gamma$.

\section{Long-distance contributions to the $B\to \rho\gamma$ decay}

An important class of weak radiative decays are those mediated by
the penguin mechanism $b\to s(d)\gamma$
\begin{equation}\label{penguin}
{\cal H}_{\rm eff} = \frac{-4G_F}{\sqrt2}V_{tb} V^*_{ts} C_7
\frac{em_b}{16\pi^2}F^{\mu\nu}
\bar s\sigma_{\mu\nu} P_R b + {\cal H}_{\rm l.d.}\,,
\end{equation}
with small long-distance contributions expected from time-ordered
products of the weak nonleptonic Hamiltonian with the minimal
electromagnetic coupling (denoted as ${\cal H}_{\rm l.d.}$).

\begin{figure}[t]
\centerline{
\mbox{\epsfysize=2.0truecm \hbox{\epsfbox{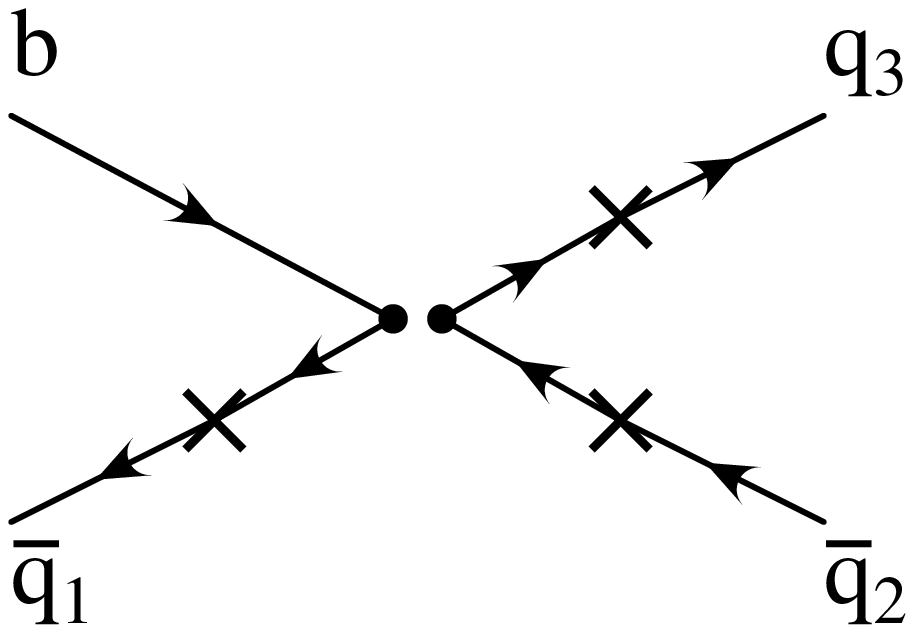}}}\quad
\mbox{\epsfysize=2.0truecm \hbox{\epsfbox{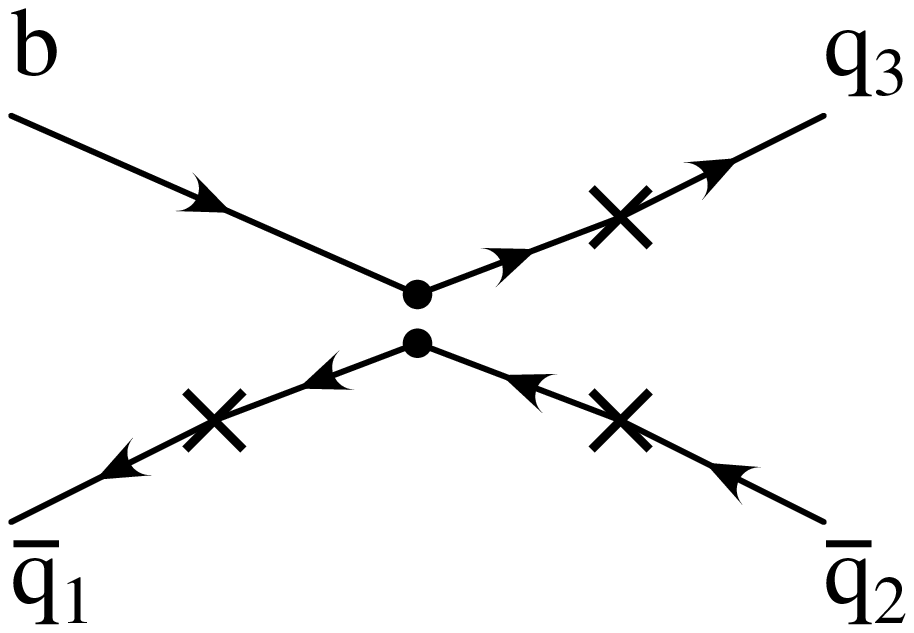}}}\quad
\mbox{\epsfysize=2.0truecm \hbox{\epsfbox{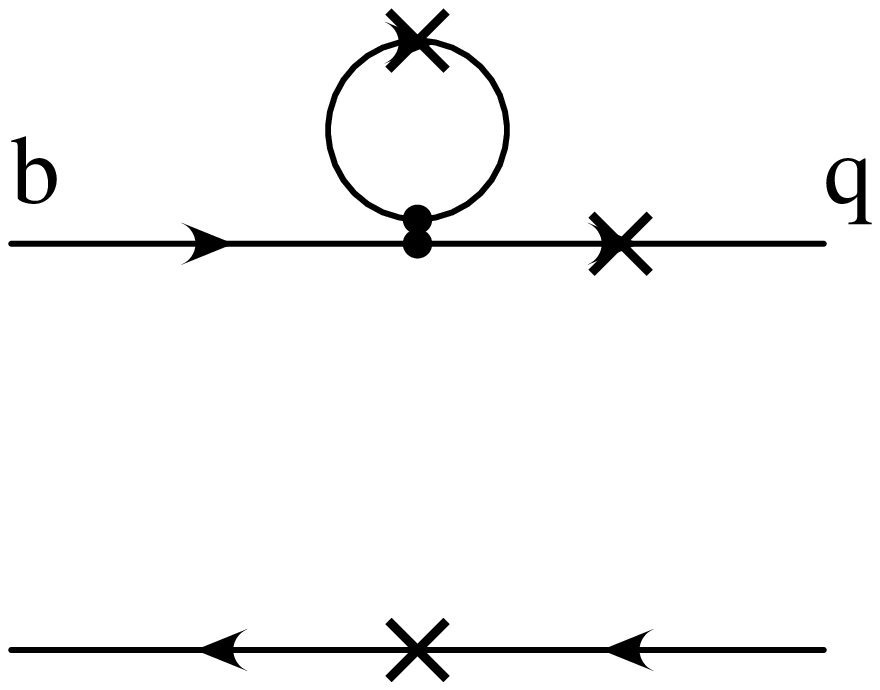}}}
}
\centerline{\mbox{(a)}\hspace{3.0cm} \mbox{(b)}
\hspace{3.0cm} \mbox{(c)}}
\centerline{
\mbox{\epsfysize=2.0truecm \hbox{\epsfbox{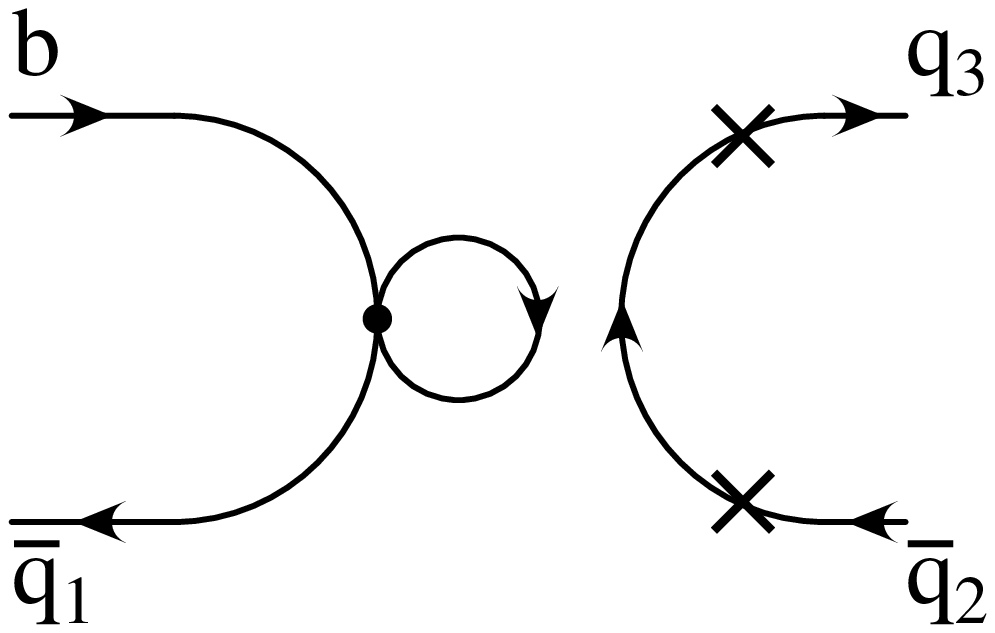}}}\quad
\mbox{\epsfysize=2.0truecm \hbox{\epsfbox{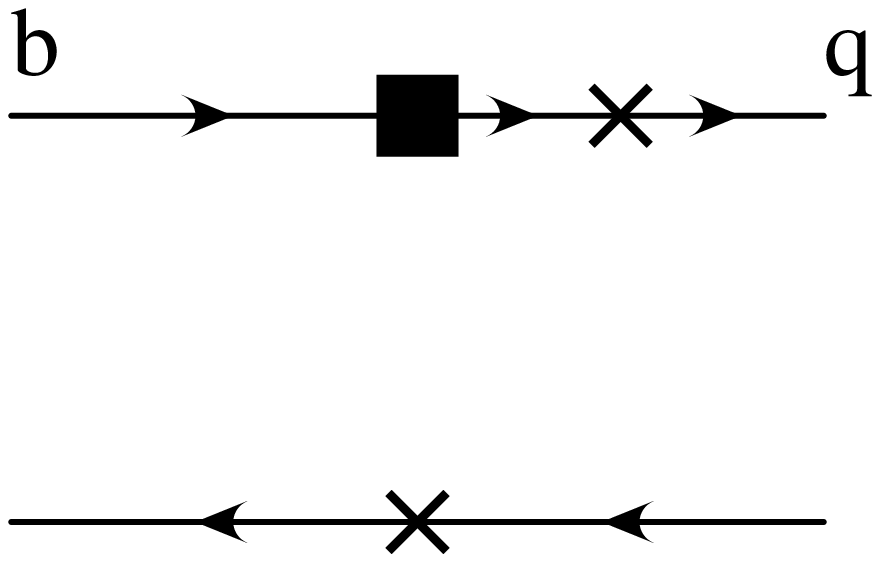}}}
}
\centerline{\mbox{(d)}\hspace{3.5cm} \mbox{(e)}}
\caption{Long distance contributions to $B\to V\gamma$ weak 
radiative decays. The crosses denote the possible attachments
of the photon to the internal quark lines. The quark diagrams
are denoted in text as a) weak annihilation $(A)$; b) $W$-exchange
$(E)$; c) penguin amplitudes containing $u$-quark ($P^{(i)}_{u}$) 
and $c$-quark ($P^{(i)}_{c}$) loops. We distinguish between 
amplitudes with the photon attaching to the spectator quark
$(i=1)$ and to the quark in the loop $(i=2)$. d) annihilation
penguin amplitudes $(PA_{u,c})$; e) gluon penguin 
amplitudes $(M^{(i)})$.
\label{fig:ld}}
\end{figure}

We show in Fig.~\ref{fig:ld} the different quark diagrams
responsible for long-distance contributions to a typical weak
radiative decay $B\to V\gamma$.
In the SU(3) limit any such amplitude can be 
written as a linear combination\cite{GP} of the amplitudes 
in Fig.~\ref{fig:ld}
with CKM factors $V_i$ (for each photon helicity $\lambda$)
\begin{equation}\label{sum}
A(B\to V\gamma_\lambda) = \sum_{q=u,c,t} V_{qb} V_{qs}^*
\sum_{{\cal M}_{i,q}=A,E,P_q,PA_q} c_{i,q} {\cal M}_{i,q\lambda}\,.
\end{equation}
This is analogous to a similar decomposition of nonleptonic
$B$ decay amplitudes into graphic amplitudes\cite{GHLR}, and
is equivalent to a more usual SU(3) analysis in terms
of reduced matrix elements. 

The long-distance amplitudes appearing in (\ref{sum}) are notoriously
difficult to calculate. They have been estimated using various
methods such as QCD sum rules\cite{KSW}, vector meson 
dominance\cite{vmd}, perturbative QCD\cite{pqcd} and Regge
methods\cite{regge}. A few sample results are tabulated in Table 1,
separately for the two helicities of the photon $\lambda=L,R$. These
calculations show that the dominant long-distance amplitude 
in $b\to d\gamma$ decays comes from the weak annihilation graph $A$
in Fig.~\ref{fig:ld}(a). 

It is therefore rather fortunate that the weak annihilation amplitude 
can be computed in an esentially model-independent way. In the 
factorization approximation, one finds\cite{GP}
\begin{equation}\label{ALR}
A_{L,R} = -\frac{G_F}{\sqrt2}(C_2 +\frac{C_1}{N_c})
em_\rho f_\rho \left( f_B + E_\gamma (f_V \pm f_A)\right)
\end{equation}
with $f_{V,A}(E_\gamma)$ the radiative leptonic form factors defined
in (\ref{fV}), (\ref{fA}), and $C_1(m_b)=-0.29$, $C_1(m_b)=1.13$
are Wilson
coefficients in the weak nonleptonic Hamiltonian. Nonfactorizable
corrections to this result arise from hard gluons connecting the
initial and final state quarks, but appear only at higher twist\cite{GP}.
Furthermore, to leading order of a twist expansion for the 
radiative leptonic formfactors $f_{V,A}$ (\ref{fVA}), the weak 
annihilation
amplitude couples predominantly to left-handed photons.
A similar suppression of the right-handed helicity amplitudes is
noted for all other long-distance contributions (see Table 1),
which has implications for proposals to search for new physics
through photon helicity effects in $b\to s\gamma$ decays\cite{newp}.

\begin{table}[t]
\caption{Estimates of the short-distance and long-distance 
amplitudes in $B\to \rho\gamma$ decays 
(in units of $10^{-6}$ GeV).
The estimates of the $WA$ and $W$-exchange amplitudes $A_\lambda$ 
and $E_\lambda$ used $R=2.5$ GeV$^{-1}$. The penguin type amplitudes
$P_u$ and $P_c$ have been estimated using vector meson
dominance.\label{Table1}}
\begin{center}
\begin{tabular}{|c|c|c|c|c|c|}
\hline
Photon helicity & $|P_{t\lambda}|$ & $|P_{c\lambda}|$ & 
$|P_{u\lambda}|$ & $|A_\lambda|$ & $|E_\lambda|$\\
\hline
\hline
$\lambda=L$ & $1.8$ & $0.16$ & $0.03$ &
  $ 0.6$ & $ 0.05$ \\
$\lambda=R$ & $0$ & $0.04$ & $0.007$ & $0.07$ & $0.007$ \\
\hline
\end{tabular}
\end{center}
\end{table}

The main motivation for measuring exclusive weak $B$ radiative decays
is connected with the possibility of extracting the CKM matrix 
element $|V_{td}|$.
The weak annihilation amplitude $A$ introduces the most significant
theoretical uncertainty in such a determination. Keeping only the
leading long-distance amplitude, one finds for the ratio of
charge-averaged rates
\begin{equation}
\frac{{\cal B}(B^\pm \to \rho^\pm\gamma)}{{\cal B}(B^\pm \to
K^{*\pm}\gamma)} = \left|\frac{V_{td}}{V_{ts}}\right|^2 R_{SU(3)}
\left(1 -
\varepsilon_A\left|\frac{V_{ub}V_{ud}^*}{V_{td}V_{tb}^*}\right|
 \cos\alpha \cos\phi_A + {\cal O}(\varepsilon_A^2)\right)
\end{equation}
with $\varepsilon_A e^{i\phi_A} = A_L/P_{tL}$ and $R_{SU(3)}=0.76\pm
0.22$ an SU(3) breaking parameter in the short-distance
amplitude\cite{LCSR}.
Using the estimates of Table 1 one finds $\varepsilon_A\simeq 0.3$,
which gives an uncertainty of about 15\% in this determination of
$|V_{td}|$ from long-distance effects. The amplitude $A$ induces also 
a  direct CP asymmetry in charged $B$ decay $A_{CP}=
2\left|\frac{V_{ub}V_{ud}^*}{V_{tb}V_{td}^*}\right|\varepsilon_A 
\sin\alpha \sin\phi_A$, which can be as large as 30\% for optimal 
values of the weak and strong phases $\alpha,\phi_A$.

\subsection{Constraining the CKM matrix with
exclusive weak radiative $B$ decays}\label{subsec:wpp}

Interference between the e.m. penguin and long-distance amplitudes
can produce isospin breaking effects in $B^\pm\to \rho\gamma$ 
decays. Some of the latter contribute with a different weak phase 
than the former, which led to the suggestion to use isospin breaking 
in these decays in order to extract information about CKM 
parameters\cite{AliBr}.

A more careful treatment\cite{bound} shows that such
an approach could receive contaminations from additional 
long-distance effects with the {\em same} weak phase as the 
short-distance amplitude.
These amplitudes can be related by SU(3) symmetry to
isospin-breaking effects in $B\to K^*\gamma$ decays.
Experimental data on these modes have been recently reported by
the CLEO\cite{CLEO}, BaBar\cite{babar} and Belle\cite{belle}
collaborations, which find
\begin{eqnarray}\label{data1}
{\cal B}(B^\pm \to K^{*\pm}\gamma) &=& (3.76^{+0.89}_{-0.83}\pm 0.28)
\times 10^{-5}\qquad (CLEO)\\
& & (2.87\pm 1.20^{+0.55}_{-0.40})\times 10^{-5}\qquad
(BELLE)\nonumber\\
\label{data2}
{\cal B}(B^0 \to K^{*0}\gamma) &=& (4.55^{+0.72}_{-0.68}\pm 0.34)
\times 10^{-5}\qquad (CLEO)\\
& &(4.94\pm 0.93^{+0.55}_{-0.52})\times 10^{-5}\qquad (BELLE)
\nonumber\\
& & (5.2\pm 0.82\pm 0.47)\times 10^{-5}\qquad (BABAR)\nonumber
\end{eqnarray}
The long-distance amplitudes responsible for the difference in rate 
between charged and neutral $B\to K^*\gamma$ modes are shown in 
Figs.~\ref{fig:ld}(c), (e), and contain a charm loop or a gluon penguin 
in which the photon attaches to the spectator quark.
\begin{figure}[t]
\centerline{
\mbox{\epsfysize=7.0truecm \hbox{\epsfbox{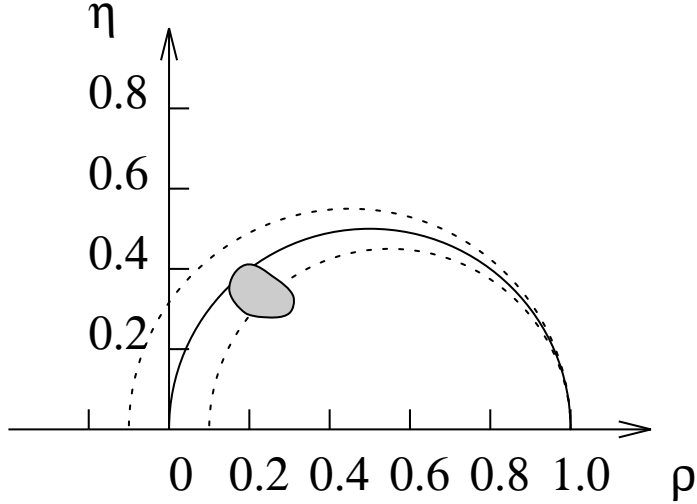}}}}
\caption{Illustrative example of a constraint on the CKM parameters 
from isospin
breaking in $B\to\rho(\omega)\gamma$ decays. The shaded area is the
68\% CL contour obtained from a global fit of the unitarity
triangle$^{26}$. The region contained between the dotted lines is the 
exclusion region corresponding to $|R_\rho-1|/\varepsilon_A = 0.1$.
\label{fig:ckm}}
\end{figure}
The preliminary data (\ref{data1}), (\ref{data2}) indicate that these
effects could be significant. It has been proposed\cite{bound}
therefore to 
eliminate them by combining $B\to \rho\gamma$ with $B\to K^*\gamma$ 
data, by forming the combined ratio
\begin{eqnarray}\label{Rrho}
R_\rho &\equiv& \frac{{\cal B}(B^\pm \to \rho^\pm\gamma)}
{2{\cal B}(B^0 \to \rho^0\gamma)}\cdot 
\frac{{\cal B}(B^0 \to K^{*0}\gamma)}
{{\cal B}(B^\pm \to K^{*\pm}\gamma)}\\
&=& 1 - 2\varepsilon_{PA_c} -
2 \varepsilon_A
\left|\frac{V_{ub}V_{ud}^*}{V_{tb}V_{td}^*}\right|\cos\alpha
\cos\phi_A + {\cal O}(\varepsilon_i^2)\,.\nonumber
\end{eqnarray}
We expanded here to linear order in the ratios of long-/short-distance
amplitudes $\varepsilon_i$. The residual contamination from 
the (OZI-suppressed) penguin-annihilation amplitude 
$\varepsilon_{PA_c}$ Fig.~\ref{fig:ld}(d) can be expected to be very small.
An upper bound on its size can be given in terms of experimental data 
on $B_s$ decays as $|\varepsilon_{PA_c}|^2 \leq 
2\Gamma(B_s\to \rho^0\gamma)/\Gamma(B^\pm\to K^{*\pm}\gamma)$.

Any measurement of the ratio (\ref{Rrho}) different from 1 can be translated
into a constraint on the CKM factors in the last term. The cleanest 
approach involves extracting the factor 
$\left|\frac{V_{ub}V_{ud}^*}{V_{tb}V_{td}^*}\right|\varepsilon_A$
from a combination of data in $B\to\gamma\ell \nu_\ell$ and $B\to
K^*\gamma$ decays\cite{bound}. This results into a bound on
the weak phase $\alpha$ which excludes values around $\alpha=90^\circ$. 

A simpler (although less model-independent) approach would
combine theoretical calculations of $\varepsilon_A$ using (\ref{ALR}),
in order to constrain the combination of CKM parameters
$\left|\frac{V_{ub}V_{ud}^*}{V_{tb}V_{td}^*}\right|\cos\alpha$.
A typical region in the $(\rho,\eta)$ plane which can be excluded in
this way is shown in Fig.~\ref{fig:ckm}. This largely
overlaps with the area presently favored by global fits of the 
unitarity triangle, and can therefore be expected to be useful in
further constraining it, as more data on weak radiative $B$ decays 
become available.

\section*{Acknowledgments}
It has been a pleasure collaborating with Ben Grinstein, Yuval Grossman, 
G. Korchemsky and Tung-Mow Yan on the issues discussed here. I am grateful 
to the organizers of the PPP workshop for the invitation to give a talk and
to the Center for Theoretical Sciences, R. O. C. for financial support.
This research was supported by the DOE grant DOE-FG03-97ER40546.


\begin{thebibliography}{99}
\bibitem{ope}
Another class of decays which can be described 
systematically corresponds to diagrams with one far offshell
photon $q^2 \gg \Lambda^2$ emitted from an internal quark line. 
The internal quark propagator is contracted to a point and
the decay amplitudes are expanded in powers of $1/q^2$ with the help
of an OPE\cite{operefs}.
\end{thebibliography}
\end{document}